# SANS contrast variation study of magnetoferritin structure at various iron loading

Lucia Melnikova[a], Viktor I. Petrenko[b,c], Mikhail V. Avdeev[b], Oleksandr I. Ivankov[b,c,d], Leonid A. Bulavin[c], Vasil M. Garamus[e], László Almásy[f], Zuzana Mitroova[a], Peter Kopcansky[a]

[a]*Institute of Experimental Physics, SAS, Watsonova 47, 040 01 Kosice, Slovakia*
[b]*Joint Institute for Nuclear Research, Joliot-Curie 6, 141980 Dubna, Moscow region, Russia*
[c]*Kyiv Taras Shevchenko National University, Volodymyrska Street 64, Kyiv, 01033 Ukraine*
[d]*Laboratory for Advanced Studies of Membrane Proteins, Moscow Institute of Physics and Technology, 141700 Dolgoprudniy, Russia*
[e]*Helmholtz-Zentrum Geesthacht: Centre for Materials and Coastal Research, Max-Planck-Street 1, 21502 Geesthacht, Germany*
[f]*Wigner Research Centre for Physics, HAS, H-1525 Budapest, POB 49, Hungary*

Corresponding author: melnikova@saske.sk, tel.: +421 55 792 2233, Fax: +421 55 633 62 92

Keywords: SANS contrast variation method, magnetoferritin, iron oxides, apoferritin

A b s t r a c t

Magnetoferritin, a synthetic derivate of iron storage protein – ferritin, has been synthesized with different iron oxide loading values. Small-angle neutron scattering experiments were applied to study the structure of magnetoferritin solutions using contrast variation method by varying the light to heavy water ratio of the solvent. Higher iron loading leads to increasing of the neutron scattering length density of magnetoferritin and also to the increase of the polydispersity of complexes. The formation of the magnetic core and the variation of the protein shell structure upon iron loading are concluded.

## 1. Introduction

In 1992 Meldrum and co-workers prepared new magnetic material called magnetoferritin. Magnetoferritin differs from the natural iron storage protein, ferritin, in the iron core composition of magnetic iron oxide phase ($Fe_3O_4$, $\gamma-Fe_2O_3$) surrounded by the protein shell – apoferritin with an outer diameter of about 12 nm [1]. Physico-chemical properties of magnetoferritin have been intensively studied. Bulte et al. studied magnetic properties of magnetoferritin and revealed superparamagnetic behavior of magnetoferritin dispersions [2]. Later, higher ratio of maghemite inside magnetoferritin was detected using Mössbauer spectroscopy in magnetic field at 9 T and at temperature 4.2 K [3,4]. The study of Wong et al.

confirmed the presence of superparamagnetic nanoparticles of $Fe_3O_4$ and $\gamma$-$Fe_2O_3$ in magnetoferritin, which was prepared by a chemical process using a controlled amount of oxidant, trimethylamine N-oxide, in anaerobic conditions, at temperature 65°C and pH 8.6. The new chemical method allowed to obtain magnetoferritin with various iron atoms per one apoferritin biomacromolecule (denoted as loading factor, LF). By transmission electron microscopy the aggregation of magnetoferritin with LF > 1000 was observed. Results have shown that with higher LF the diameter of nanoparticles increases [5,6]. The effect of dipolar interaction on temperature relaxation of magnetoferritin was studied and results were in a good agreement with theoretical models [7]. Kasyutich et al. in 2008 prepared three-dimensional arranged crystals of magnetoferritin with size about 10 to 100 micrometers by protein crystallization technology. [8,9]. Detailed investigation of magnetoferritin nanoparticles in solution using transmission electron microscopy, X-ray diffraction, atomic force microscopy by Martínez-Pérez et al. have shown that the external diameter of magnetoferritin nanoparticles increased with LF. The reason of this effect was related to conformational changes of the protein due to binding of iron oxides [10]. Recent studies of magnetically induced optical birefringence for colloidal dispersion of magnetoferritin in comparison with ferritin have shown differences in their magneto-optical behavior [11,12]. Magneto-optical Faraday rotation at room temperature in applied magnetic field with intensity H = 2970 Oe (i.e. 236.4 kA.m$^{-1}$) showed different spectral dependences of the Faraday rotation on the wavelength allowing to distinguish ferritin and magnetoferritin. Dependence of the Faraday rotation of magnetoferritin as a function of applied magnetic field was characteristic for superparamagnetic system and could be modeled by the Langevin function with a log-normal size distribution of the magnetic particles [13]. Magnetoferritin was found to decompose hydrogen peroxide in the presence of the substrate, N,N-diethyl-p-phenylenediamine sulfate, which changed the color after the reaction. The peroxidase-like activity of magnetoferritin was more pronounced in the case of higher LF [14]. Physico-chemical characterization of magnetoferritin biomacromolecules in terms of morphology, structural and magnetic properties shows that iron oxides loaded into apoferritin affects the structure of protein shell [15,16]. However, the specific effect of magnetic nanoparticles and the role of the different amount of iron inside one biomacromolecule on the protein structure have not been fully determined yet.

Extensive studies of magnetoferritin from the first synthesis in 1992 showed that its biological origin, well-defined diameter of nanoparticles, colloidal stability and superparamagnetic behavior provide significant potential of magnetoferritin for application in nanotechnology, industry, but especially in the biomedical sciences and cell biology [1,17]. The ability of magnetoferritin to undergo structure change under certain conditions enhances their potential for various applications. Magnetoferritin structure can be modified by other substances [18], drugs, surfactants, signal molecules, antibodies [19,20], resulting in change of the protein shell, so that the molecule can be completely closed, or partially open, or completely disrupted. Various potential applications are reported through chemical modifications of the external or internal surface, such as in magnetic resonance imaging of tissues [2,21], as contrast agent for cell labeling [22], as standard for diagnosis of various diseases [11-13], in nanocatalytical chemistry [14], cell separation [20] or in targeted transport of drugs. The drug binding to magnetoferritin allows visualization of pathological tissues or targeted transport directly only to the damaged area of the organism without the side effects of the drug on healthy tissues and organs [23]. All these applications rely on the ability of magnetoferritin to pertain or to change its core-shell structure.

Small-angle neutron scattering (SANS) is an experimental method particularly suited to study the details of such core shell type objects of sizes 10-20 nanometers. Contrast variation technique in SANS can be used to reveal fine details of dispersions of multicomponent particles and clusters in liquid media. The structure and interactions of magnetic or non-magnetic particles can be obtained by the polarized neutron scattering analysis. A versatile approach to study polydisperse and nonhomogeneous particle dispersions by using contrast variation has been recently introduced [24,25]. Modifications of classical expressions describing changes in integral parameters of scattering (forward scattering intensity, radius of gyration, Porod integral) with the contrast are analyzed and the shape and size polydispersity is calculated. [24,25].

In the present work, we investigated the structural changes of magnetoferritin with various loading factors by SANS contrast variation method. Results have shown the increase of effective match points and polydispersity of magnetoferritin with the LF growth. The magnetic core in magnetoferritin shifted the effective match points in comparison with pure apoferritin. Non-zero value of the forward scattering intensity in the effective match points indicated structural polydispersity, i.e. the non-uniform distribution of the iron oxide inside the protein cage.

## 2. Materials and methods

### 2.1 Chemicals

Ammonium ferrous sulfate hexahydrate $((NH_4)_2Fe(SO_4)_2.6H_2O)$, equine spleen apoferritin in 0.15 M NaCl, 3-[(1,1-dimethyl-2-hydroxy-ethyl)amino]-2-hydroxylpropane-sulfonic acid (AMPSO), hydrogen peroxide $(H_2O_2)$, hydrochlorid acid (HCl), potassium thiocyanate (KSCN), sodium hydroxide (NaOH) and trimethylamine N-oxide $(Me_3NO)$ were obtained from Sigma-Aldrich.

## 2.2 Magnetoferritin preparation

Magnetoferritin was synthesized by incorporation of ferrous ions into the apoferritin hollow cage in 0.05 M AMPSO, pH 8.6 and anaerobic conditions at 65°C. Magnetoferritin complexes with four loading factors of 160, 510, 766 and 910, were prepared. The LF was determined spectrophotometrically using UV-VIS spectrophotometer SPECORD 40, Analytik Jena, with precision about 1%. Concentration of iron was measured after $HCl/H_2O_2$ oxidation and KSCN addition by light absorption of thiocyanate complex at wavelength ($\lambda$) 400 nm. Protein amount was detected using standard Bradford method at $\lambda = 595$ nm. After the synthesis samples were 24 h freeze dried to obtain a powder. Finally, 10 mg/ml solutions of protein were prepared by dissolving powders in AMPSO alkaline buffer solution with various ratio of $H_2O:D_2O$ for contrast variation method. AMPSO alkaline buffer solutions of similar $H_2O:D_2O$ ratios were used for incoherent background measurements.

## 2.3 Small-angle neutron scattering experiment

SANS measurements were carried out on small-angle neutron scattering spectrometer YuMO [26,27] installed on the pulsed nuclear reactor IBR-2 at the Joint Institute for Nuclear Research (JINR, Russia). Samples were located in standard quartz cuvettes (Hellma) with a thickness of 1 mm and the measurements were performed at room temperature in the absence of external magnetic field. Thermal non-polarized neutrons of wavelengths between 0.5 Å and 8 Å were used. The measured scattering curves were corrected for the scattering from the solvent and the absolute cross section was determined by vanadium calibration standards.

## 3. Results and discussions

The small-angle neutron scattering from magnetoferritin aqueous suspensions was measured at the different $D_2O$ content in the solvent. Experimental scattering curves for apoferritin (concentration about 12 mg/ml) and magnetoferritin with LF 160, 510, 770, and 910 in 100% $D_2O$, after subtraction of scattering from AMPSO buffer, are shown in Fig. 1. Smoothing of the measured SANS curve in comparison with pure apoferritin is seen like in previous works [15,16]. Similar behavior of SANS curves at varying $H_2O/D_2O$ content is obtained for all studied LFs except for LF 910. For magnetoferritin with LF 910 dark precipitates sedimented to the bottom of the cell. Therefore, only MFer 910 in 100% $D_2O$ was measured. As an example, SANS curves of magnetoferritin with LF 770 in various ratios of $H_2O$ and $D_2O$ with protein concentration of 10 mg/ml are shown in Fig 2a.). From linear fitting of Guinier plots (Fig. 2b.)) values of the forward scattering intensity, I(0), and radius of the gyration, $R_g$ were obtained. The scattering length densities (SLDs) of the solvent mixtures, $\rho_s$, were calculated using the SLD of H$_2$O and D$_2$O values ($\rho_{H2O}$ = -0.559 · 10$^{10}$ cm$^{-2}$, $\rho_{D2O}$ = 6.34 · 10$^{10}$ cm$^{-2}$). The lowest value of I(0) corresponds to the match point. For homogeneous particles, the scattering length density of the solvent and particles coincides in the match point. Non-zero value of the forward scattering intensity in the match point, as shown in Fig. 3, indicates structural polydispersity in the system. The match point is shifted with increasing loading factor (Fig. 3). In this analysis, the magnetic scattering intensity contribution can be neglected. The contribution of magnetic scattering for LF 770 is less than 5%, assuming magnetite in the protein cage. Magnetometry data on similar samples indicated magnetization values below 0.05 A m$^2$ kg$^{-1}$ [6,14], which is much lower than the specific magnetization of apoferritin filled with magnetite phase. Already in the monodisperse approximation and assuming that the magnetic core in magnetoferritin consists of magnetite, the shifts of the effective match points as compared to the protein moiety of apoferritin (SLD 2.34 · 10$^{10}$ cm$^{-2}$) give 0.026, 0.099 and 0.114 respectively, for the volume fractions of magnetic material in the magnetoferritin cage (corresponding SLDs are 2.46 · 10$^{10}$ cm$^{-2}$, 2.79 · 10$^{10}$ cm$^{-2}$, and 2.86 · 10$^{10}$ cm$^{-2}$ for LF 160, LF 510 and LF 770, respectively). These apparent volume fractions obtained from the measurements are about 3-5 times larger than what could be caused by the iron oxide loading in the synthesis (0.005, 0.017 and 0.025 for LF 160, LF 510 and LF 770, respectively). The calculated match points are significantly higher than those for native ferritin with different amount of iron [28]; they correspond to native ferritin with about 800 and 2000 iron atoms per protein shell. Thus, SANS contrast variation measurements indicated an abnormally high ratio of the amount of the iron oxide phase to the protein. The reason of this increase can be explained by the partial destruction of the shell, and thus the effective growth of the relative content of the magnetic component in the magnetoferritin. The residual scattering in the effective match points increases in accordance with the broadening of the polydispersity function for larger LFs, in agreement with the increased polydispersity observed in magnetooptical measurements [11,12]

The precise mechanism of magnetic loading influence on the protein structure is not known yet. In a recent study apoferritin disassembly was observed at pH below 3.4 [29]. This is however not our case, since our magnetoferritin was prepared in alkaline pH (pH 8.6 for each synthesis). While it was not possible to control pH directly during the synthesis process due to the hermetically enclosed reaction bottle, still after the synthesis the pH value was checked and only slight pH change, namely some decrease for higher LFs, was detected but the minimum pH after the synthesis was above 7 for all LFs.

The observed disassembly of magnetoferritin cage can be related to the influence of the magnetic nanoparticles. Interaction between magnetic nanoparticles and proteins is the subject of extensive studies in recent years and the general conclusion is that different nanoparticles with specific concentration, surface and size can affect proteins structural conformations [30-34].

**4. Conclusions**

We studied the structural changes of magnetoferritin in aqueous solution with LFs: 160, 510, 770 and 910 by contrast variation small angle neutron scattering. The obtained values of the average neutron scattering length density of magnetoferritin (effective match point) and its variation among the aggregates increased with iron loading. This effect can be associated with formation of magnetic iron oxide nanoparticles inside complexes and their influence on the protein structure. While the scattering data could give only an average information on the structural properties of the complexes, other experimental techniques, such as electron microscopy or fluorescence spectroscopy could serve with more direct information on the structure and conformational changes of magnetoferritin upon variation of the iron loading.

**Acknowledgements**

This work was supported by the Slovak Scientific Grant Agency VEGA (projects No. 0041, 0045), by the European Structural Funds (PROMATECH), projects NANOKOP No. 26110230061 and 26220120021, PHYSNET No. 26110230097, APVV 0171–10 (METAMYLC) and M-ERA.NET MACOSYS.

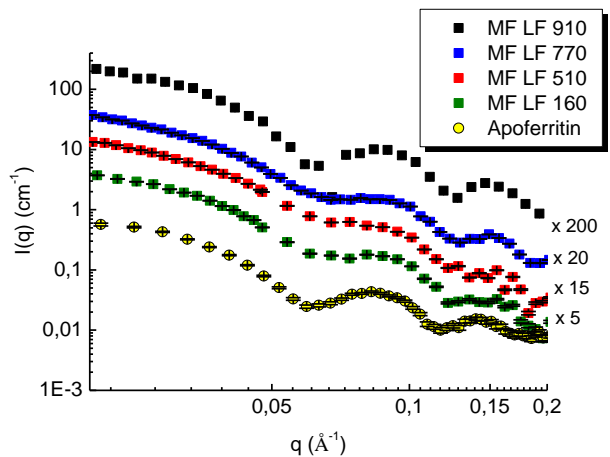

Fig. 1 Experimental scattering curves of magnetoferritins (LF 160, 510, 770, 910) and apoferritin in 100% $D_2O$ with protein concentration of about 12 mg/ml for apoferritin and 10 mg/ml for magnetoferritin samples.

a.)                                                  b.)

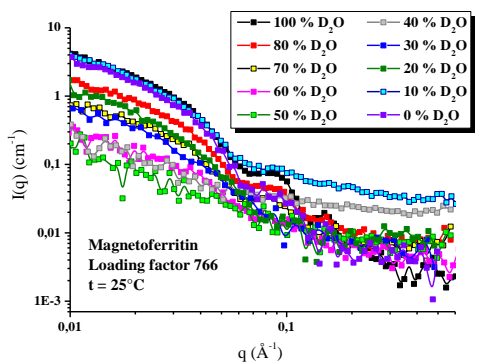 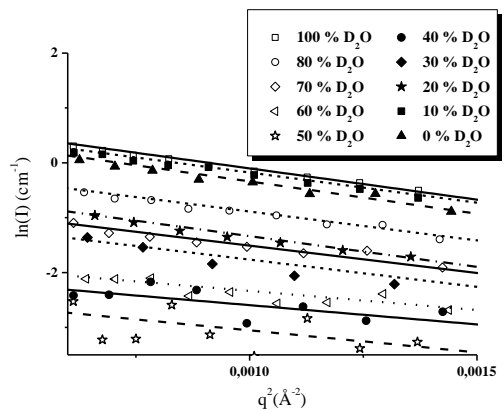

Fig. 2 Scattering curves of magnetoferritin LF 770 in various ratios of $H_2O$ and $D_2O$ with protein concentration of 10 mg/ml (a) and the corresponding Guinier plots (b).

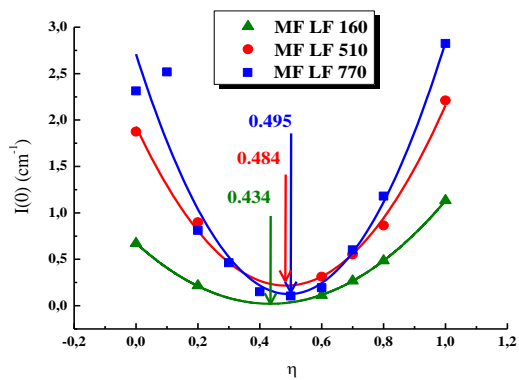

Fig. 3 Scattering intensity at the zero angle as a function of the $D_2O$ content ($\eta$). The minima of the fitted parabolas show the effective match points at the studied loading factors.